# Overhaul and Installation of the ICARUS-T600 Liquid Argon TPC Electronics for the FNAL Short Baseline Neutrino Program

L. Bagby[a], B. Baibussinov[b], B. Behera[m], V. Bellini[c], R. Benocci[d], M. Betancourt[a],
M. Bettini[b], M. Bonesini[d], T. Boone[m], A. Braggiotti[b,e], J. D. Brown[a], H. Budd[n], F. Calaon[b],
L. Castellani[b], S. Centro[b], A. G. Cocco[g], M. Convery[l], F. Fabris[b], A. Falcone[d], C. Farnese[b],
A. Fava[a], F. Fichera[c], M. Giarin[b], D. Gibin[b], A. Guglielmi[b], R. Guida[b], C. Hilgenberg[m],
B. Howard[a], W. Ketchum[a], S. Marchini[b], A. Menegolli[f], G. Meng[b,1], C. Montanari[a,2],
M. Mooney[m], G. Moreno Granados[p], J. Mueller[m], M. Nessi[i], M. Nicoletto[b], R. Pedrotta[b],
R. Peghin[b], G. Petrillo[l], F. Pietropaolo[i,3], G. Rampazzo[b], A. Rappoldi[f], G. L. Raselli[f],
M. Rossella[f], C. Rubbia[h,i,j,4], A. Scaramelli[f], F. Sergiampietri[k], M. Spanu[o], D. Torretta[a],
M. Torti[d], F. Tortorici[c], Y. T. Tsai[l], M. Turcato[b], F. Varanini[b], S. Ventura[b], F. Vercellati[f],
C. Vignoli[h], D. Warner[m], R. J. Wilson[m], M. Worcester[o], A. Zani[i], P. G. Zatti[b] for the
ICARUS Collaboration

[a] *Fermi National Accelerator Laboratory, Batavia, Illinois, USA*
[b] *Dipartimento di Fisica e Astronomia "G. Galilei", Università di Padova and INFN, Padova, Italy*
[c] *Dipartimento di Fisica e Astronomia, Università di Catania and INFN, Catania, Italy*
[d] *Dipartimento di Fisica "G. Occhialini", Università di Milano-Bicocca and INFN Milano-Bicocca, Milano, Italy*
[e] *Istituto di Neuroscienze, CNR, Padova, Italy*
[f] *Dipartimento di Fisica, Università di Pavia and INFN, Pavia, Italy*
[g] *Dipartimento di Scienze Fisiche, Università Federico II di Napoli and INFN, Napoli, Italy*
[h] *INFN - Laboratori Nazionali del Gran Sasso, Assergi, Italy*
[i] *CERN, Geneva, Switzerland*
[j] *GSSI, L'Aquila, Italy*
[k] *INFN, Pisa, Italy*
[l] *SLAC National Accelerator Laboratory, Stanford, USA*
[m] *Colorado State University, Fort Collins, Colorado, USA*
[n] *University of Rochester, Rochester, New York, USA*
[o] *Brookhaven National Laboratory, Upton, New York, USA*
[p] *Centro de investigacion y de Estudios Avanzados del IPN (Cinvestav), Mexico City, Mexico*

E-mail: guang.meng@pd.infn.it

---

[1] Corresponding Author
[2] On leave of absence from INFN Pavia
[3] On leave of absence from INFN Padova
[4] Spokesperson

ABSTRACT: The ICARUS T600 liquid argon (LAr) time projection chamber (TPC) underwent a major overhaul at CERN in 2016-2017 to prepare for the operation at FNAL in the Short Baseline Neutrino (SBN) program. This included a major upgrade of the photo-multiplier system and of the TPC wire read-out electronics. The full TPC wire read-out electronics together with the new wire biasing and interconnection scheme are described. The design of a new signal feed-through flange is also a fundamental piece of this overhaul whose major feature is the integration of all electronics components onto the signal flange. Initial functionality tests of the full TPC electronics chain installed in the T600 detector at FNAL are also described.

# Contents

## 1. Introduction

The liquid argon time projection chamber (LAr-TPC) imaging detection technique [1] which allows for accurately identifying and reconstructing ionizing tracks in complex particle interactions, has been brought to full maturity with the successful operation of the liquid argon large mass T600 detector at the LNGS underground laboratories where it was exposed to CNGS beam and to cosmic rays [2].

The ICARUS-T600 detector consists of a large cryostat containing two identical adjacent modules with internal dimensions 3.6 x 3.9 x 19.6 m$^3$, filled with about 760 tons of ultra-pure liquid argon continuously purified to prevent absorption of ionization electrons by electronegative impurities [3]. Each module houses on both vertical long sides two TPCs separated by a central, vertical, common cathode (Figure 1). A uniform electric field ($E_D$ = 500 V/cm) allows for the drift without distortion of ionization electrons produced by charged particles along their path toward the corresponding TPC. Each TPC has three parallel read-out wire planes, 3 mm apart, facing the drift volume (1.5 m deep). The wire pitch is 3 mm for all planes. The first plane (Induction-1) has horizontal wires, while the intermediate plane (Induction-2) and the third plane (Collection) have wires at +/- 60$^0$ with respect to the horizontal direction. The maximum wire length is 18 m for the Induction-1, split in two separated 9 m wires read by its electronics channels. Each TPC has a total of 13312 wires. Globally, 53248 wires are installed in the T600 (4 identical TPCs). By appropriate voltage biasing of the three wire planes full transparency of the first two planes (Induction-1 and Induction-2) might be achieved.

The two Induction planes provide bi-polar signals, while the Collection plane collects the full charge as a uni-polar signal. By combining the wire coordinate on each plane at a given drift time, a three-dimensional image of the ionizing event can be reconstructed (Figure 1). The measurement of the absolute time of the ionizing event by the scintillation light signals detected with an array of photo-multiplier tubes [4], combined with the electron drift velocity information ($v_D$ ~1.6 mm/μs at $E_D$ = 500 V/cm), provides the absolute position of the track in the drift volume with a remarkable resolution of ~1 mm$^3$. In addition, the charge signal detected in the Collection view, proportional to the deposited energy, allows for calorimetric measurement of the particle energy. A detailed description of the ICARUS detector can be found in Ref. [5].

After the successful three year run at LNGS investigating the LSND-like $\nu_\mu - \nu_e$ oscillations in the CNGS neutrino beam [6], ICARUS-T600 has been overhauled for deployment as a far detector in the SBN program at FNAL to definitively clarify the LSND effect in the Booster neutrino beam [7]. For the SBN program, we designed better performing and more compact TPC electronics, described in this paper. Furthermore, the first functionality tests on the full TPC read-out electronics chain installed in the T600 detector at FNAL are presented.

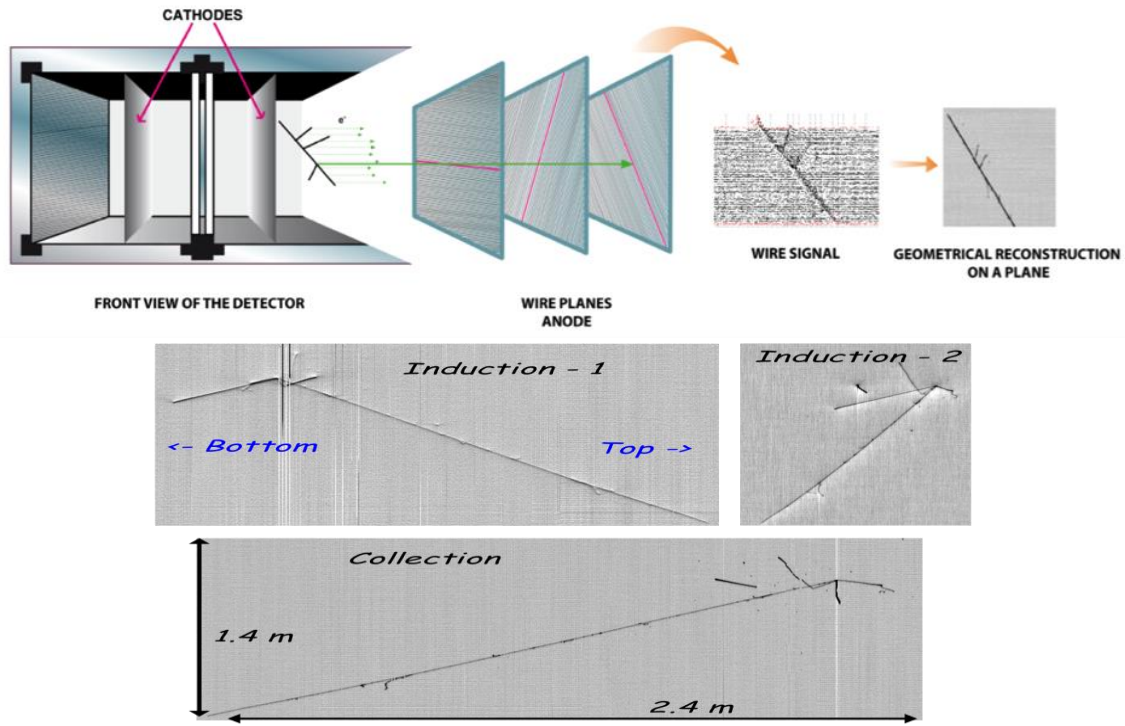

**Figure 1.** Sketch of the detector layout showing the LAr-TPC working mechanism (top). An upward-going atmospheric muon neutrino event ($E_{DEP}$ ~ 1.7 GeV) collected by ICARUS-T600 at LNGS as measured in the three - Induction-1, Induction-2 and Collection - views is shown (bottom).

## 2. The ICARUS TPC read-out electronics

### 2.1 Initial design for LNGS run

The electronics initially equipping the T600 during the LNGS operation was designed to allow for continuous read-out, digitization and independent waveform recording of signals from each wire of the TPC for the whole drift time (~1 ms) [5]. The read-out chain was organized in a 32-channel modularity. A Decoupling and Biasing Board (DBB) passed the signals received from the TPC to an Analogue Board via decoupling capacitors and provided the wire biasing. The Analogue Board hosted the front-end amplifiers, performed 16:1 channel multiplexing and 10-bit ADC digitization at 400 ns sampling time per channel. The overall gain was about 1000 electrons per ADC count, setting the collection plane signal of a minimum ionizing particle (m.i.p.), which produces ~15000 free electrons in 3 mm path, corresponding to 15 ADC counts with a dynamic range of about 70 m.i.p.

The signals from the wires were extracted from the cryostat via ultra-high vacuum feed-through proprietary flanges on top of 96 chimneys, then through twisted pair flat cables, grouped in sets of 18 cables serving 576 channels. Next to each flange on the top of the detector a rack housed the analog crate with 18 boards with amplifiers, multiplexers and 10-bit ADCs. The converted data were serialized with an industrial standard serial bus (TIA/EIA-644 LVDS) and sent to 18 digital boards in the digital crate. The digital crate was based on a VME standard while the analogue crate had a proprietary design relying on VME mechanics with a custom backplane, that allowed inserting the 32-channel analogue boards from the front and the corresponding DBB from the back. The custom designed low-noise ICARUS front-end amplifier and multiplexed ADC system used in the LNGS run performed efficiently with a signal to noise (S/N) ratio ~10, allowing collection of several thousands of neutrino and cosmic events with unprecedented imaging quality. However, the signal shaping chosen at the time with a baseline restorer presented some limitations on signals produced by the intermediate Induction 2 wire plane in case of dense showers.

The data acquisition (DAQ) of the T600 and the front-end dual channel BiCMOS circuits were conceived in 1997. The whole system was engineered and built by CAEN company (Viareggio, Italy) according to the requirements of the experiment and operated for the first time in 2001 in the INFN Pavia laboratory.

## 2.2 Upgraded design for FNAL run

The new electronics designed for the ICARUS-T600 for shallow depth operation at FNAL improves the performance of the system and drastically reduces costs and volume by using new, more advanced components. One evident limitation of the original T600 DAQ was the data rate due to the choice of the VME standard (8-10 MB/s) which is perfectly suited for underground operation at LNGS. This is a limitation for the operation of the ICARUS experiment at the FNAL Booster beam, given the shallow depth and the high rate of cosmic rays.

The new system maintains the previous architecture, which allows for a continuous triggerable multi-buffered waveform recorder for each wire of the detector with a more advanced design. In particular each channel has a dedicated serial 12-bit ADC (avoiding multiplexing), and the analog and digital parts are now integrated in one electronics module that serves 64 channels. Also, the sustainable data rate has been improved by adopting serial bus architecture with optical links for Gb/s data transmission. The proprietary flanges, that still serve 576 channels each, were also modified in order to accommodate electronics modules and DBB's directly inserted on the external and internal sides of the flanges, respectively.

The design, prototyping and performance of the new ICARUS front-end electronics chain have been described in depth in Ref. [8]. The extensive test campaign performed on dedicated LAr-TPC prototypes demonstrated the improved capability of handling efficiently the signals in the intermediate Induction-2 wire plane with a significant increase of S/N, allowing for charge measurement in this view.

In this paper, the description of the new TPC electronics follows a topological order starting from the electronic devices external to the detector and their power supply, the feed-through mechanics and eventually the decoupling circuit. The front-end amplifier and the digital board already presented in Ref. [8] are also summarized.

## 3. The front-end amplifier and the Digital Board A2795

The T600 in the SBN program will operate at shallow depth in a dedicated building. Special care has been given to the grounding systems to keep building ground and detector ground separated to preserve the noise characteristics of the TPC electronics. Ufer approach is used for the building ground. Ufer is an electrical earth grounding method using concrete-encased electrodes to improve grounding in dry areas. The detector ground is defined as the cryostats of the massive T600.

The two cryostats (cold boxes) are placed inside a warm box, isolated by elastomer layered feet. The two cryostats are electrically connected together via a heavy-duty braid, bolted to each cold box. The warm box is connected to the building Ufer structure via two 1/0 AWG cables. All building referenced cryogenic piping is isolated from the cryostats with dielectric breaks. All motors, generators, variable frequency drive equipment, and building utilities are powered from building power. All sensitive detector electronics, including the TPC and their power supplies, are referenced to the detector ground. To maintain the integrity of the 'low-noise' ground system, an impedance monitor is utilized to constantly monitor the impedance between the building ground and the detector ground. The separation between the ground structures is realized with two custom designed saturable inductors, connected to each cryostat via 1/0 AWG cables. The inductors provide an AC impedance against unwanted electrical disturbances while maintaining a low resistive DC safety ground connection to the cryostats.

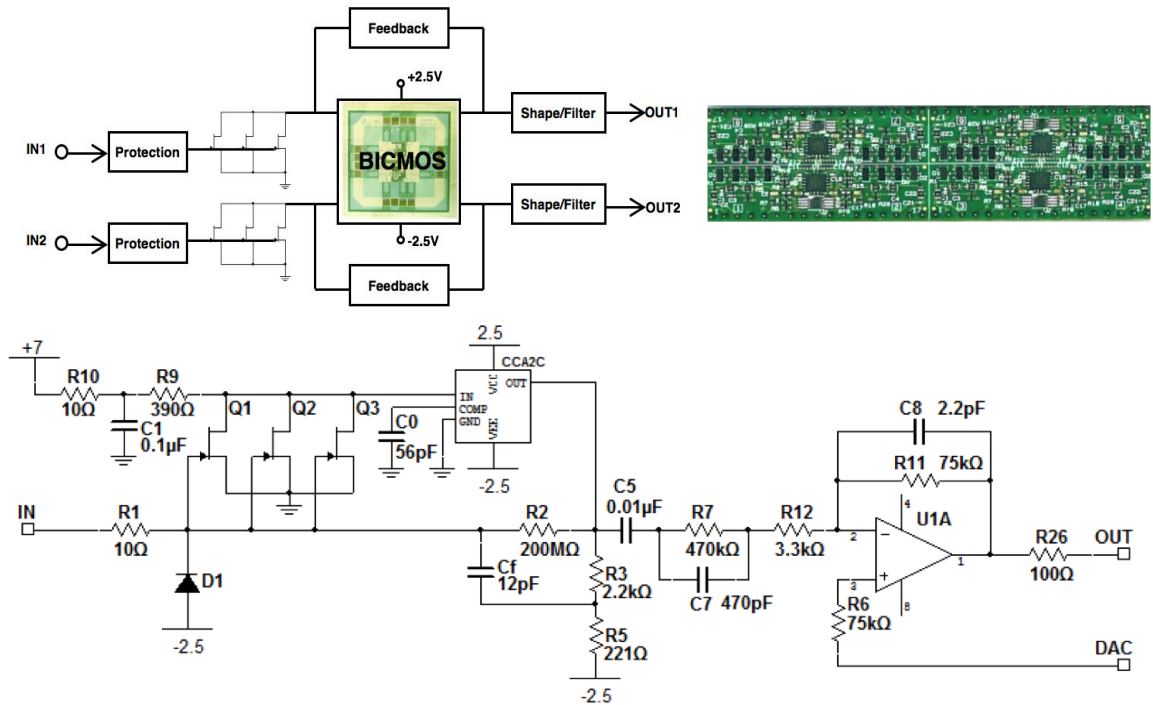

**Figure 2.** Block diagram of the front-end amplifier (top left) and PCB (81 x 20 mm$^2$) housing 8 amplifiers with a symmetrical layout (top right). Front-end schematic (bottom). The high number of elements inside the BiCMOS prevents showing a detailed schematic here.

The adopted amplifier architecture follows the ICARUS original design [5]: a Radeka-like amplifier with 3 parallel jFet as an input stage, followed by an unfolded cascode integrated in a dual channel BiCMOS ASIC, CCA2C (Figure 2). Multiple jFet (BF861C) at the input have been adopted to achieve a total transconductance $g_m \sim 45$ mS. The cascode defines the drain voltage to guarantee a constant total current through the three jFet, hence, a uniform total $g_m$ from amplifier to amplifier. The maximum total input capacitance of the amplifier due to the cable and Induction-1 wire lengths (~4-5 m, 50 pF/m and ~9 m, 20 pF/m, respectively) is of the order of 400 pF. This means the serial noise is dominant, and its effect must be mitigated by high transconductance at the input stage.

The ASIC (CCA2C) custom design was chosen for performance uniformity and compactness. To further improve it, a new even smaller custom package (4x4 mm$^2$) has been chosen. The input stage is followed by a shaper (a classic pole-zero cancellation filter) and baseline restorer with a peaking time of 0.6 μs in response of a ∂-like input current at zero detector capacitance. This choice mitigates the problem faced in case of bipolar signals as in Induction-2 channels preserving the bipolar shape and allowing for proper treatment and reconstruction even for dense showers, as demonstrated on the data collected with the test LAr-TPC [8]. The same shaping time was chosen both for Induction and Collection signals.

Eight amplifiers are mounted on a single board shown in Figure 2 with the circuit schematic. A symmetric layout was adopted to keep uniform the effects due to layout stray capacitances. Eight such small amplifier boards fit in each of the 8 connectors on the new A2795 CAEN motherboard for a total of 64 channels.

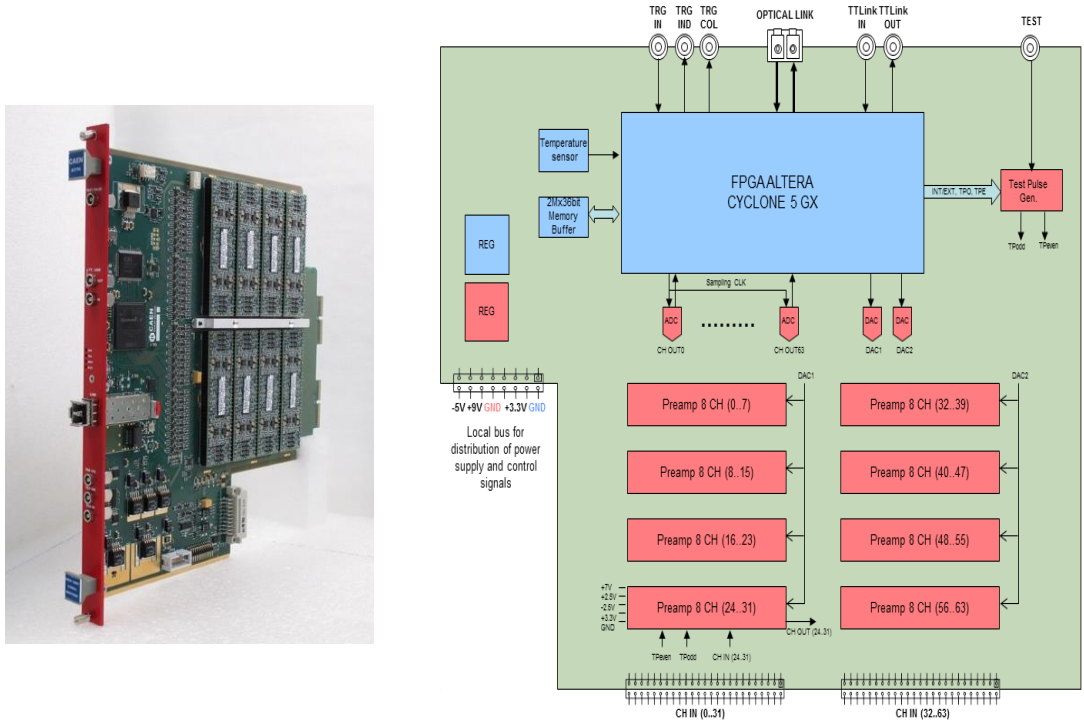

**Figure 3.** Left: A2795 custom board housing 64 amplifiers (far end), AD converter, digital control, and optical link (front panel). Right: A2795 Block Diagram.

The A2795 motherboard has been designed, engineered, and built by CAEN, in collaboration with ICARUS team according to the experimental requirements. A throughput exceeding 10 Hz has been realized by a modern switched I/O system where transactions are carried out over an optical 1.25 Gb/s serial link. After the pole-zero cancellation filter, in front of the serial ADC, a Bessel filter has been inserted. It preserves the area (that contains the charge useful information) of the filtered signal in the pass-band (roll-off frequency of 4 MHz) due to its linear phase response, and reduces the RMS serial noise. This filter interfaces each amplifier with its serial 12-bit (Least Significant Bit (LSB) = 0.8 mV, 400 ns sampling time) ADC (AD7276BUJZ) mounted onto the A2795 board. A dedicated ADC for each channel implements a synchronous conversion, eliminating the need for any multiplexing. Data buffering, digital processing, and transmission onto the optical link are performed in a programmable FPGA (Altera Cyclone 5 GX). In Figure 3, the specially shaped A2795 is shown along with its functional block diagram.

Nine A2795 boards are housed in a custom "mini-crate" mounted onto the feed-through flange designed for the transmission of the TPC wire signals (INFN proprietary design). This was an important achievement dramatically reducing the volume of the front-end electronics for each flange, serving 576 channels, from ~600 liters of the original design (a full rack) to 10 liters. The mini-crate and flange will be described in section 5.

## 4. Linear power supply

The guidelines for designing the new ICARUS Low Voltage Power Supply (LVPS) were set by the requirement of an extremely low noise (< 5.0 mV$_{pp}$) necessary for preserving the front-end performance. This implied the use of linear DC power supplies instead of a switching type, at the cost of a lower conversion efficiency (on the order of 50%) and higher weight. In addition, the LVPS system can be accessed remotely via a USB and TCP/IP interface in order to monitor voltages, currents, internal temperatures and to switch ON/OFF. Last but not least, the LVPS has to meet the "Electrical Design Standards for Electronics to be used in Experiment Apparatus at Fermilab" requirements [9].

The power compliance has been verified with preliminary measurements on a fully populated mini-crate. A sufficient margin of safety has been added for the output currents. A residual ripple < 5.0 mV$_{pp}$ has been reached. The power needed on the input side has been determined by doubling the power capability on the output of the LVPS: an AC input power of ~400 W ensures ~200 W of output. The resulting main operational parameters of LVPS are summarized in Table 1.

The most critical voltage is the +8.2V that generates the +7V with a regulator on board. This voltage defines the drain currents in the three jFet that determine the total transconductance of front-end amplifier (Figure 2) and its noise performance. All the other voltages are far less critical. However, we decided to have all generated by linear power supplies in order to avoid any noise contribution from switching frequencies.

**Table 1.** Main operational parameters and features of ICARUS LVPS main module.

| - | 115 Vac, 400 W main supply |
|---|---|
| - | 4 Low Voltage outputs: +8.2 V 18 A; +3.3 V 12 A; +12 V 1.5 A; -5.0 V 2 A |
| - | Max 1.5 Amp Fan Supply (selectable 12 V, 9 V or 7 V) |
| - | Selectable Local/Remote Control and Monitor |
| - | USB and TCP/IP interfaces |

The ICARUS LVPS module and its general architecture are shown in Figure 4 and Figure 5.

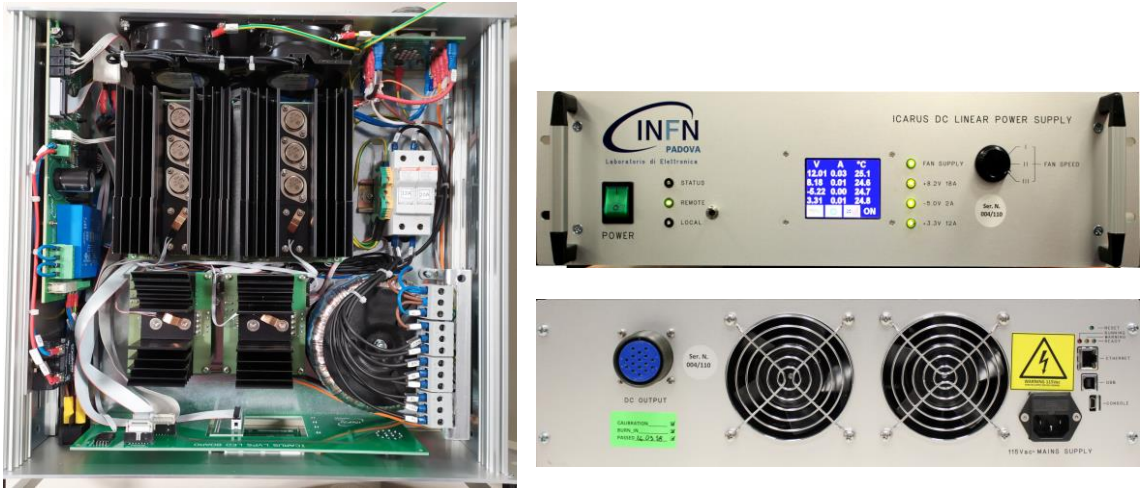

**Figure 4.** Inside (left) and front and rear panels (right) of the ICARUS LVPS module.

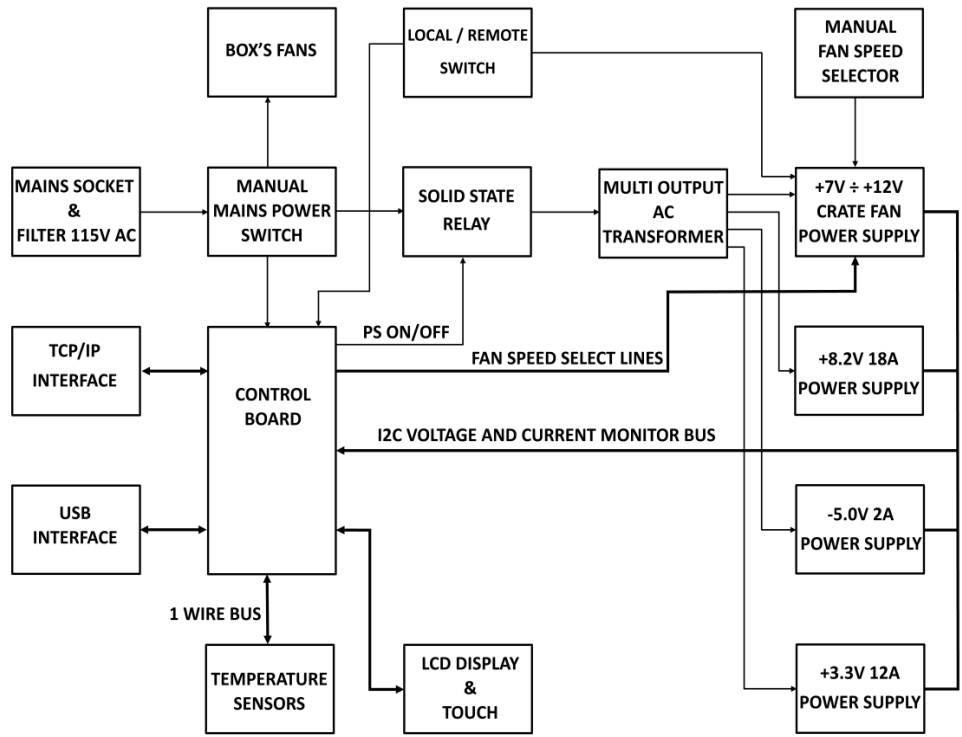

**Figure 5.** General architecture of the ICARUS LVPS module.

The LVPS main module has been designed to fit a 3U 19-inch rack with 400 mm depth. A single toroidal transformer for the power supply has been chosen. The four power supply modules contained in the LVPS main module have been engineered according to the maximum power required.

Two different basic designs have been adopted, one for the two more-demanding high-power modules and one for the less-demanding low-power module. The LVPS system is protected from over-current with fuses. One slow blow fuse protects the 115 Vac input, rated for the maximum power of the crate and able to manage the current transients. Other slow blow fuses are mounted on each secondary winding of the toroidal power transformer, dimensioned to manage both the RMS and the transient's currents. The low-power modules are protected from output over-current and overheating directly from the LDO (Low Drop Out) regulator, while the high-power modules are protected from output over-voltages and over-currents via dedicated circuits. A current-fold-back protection configuration is used against the over-current to limit the output voltage and current in case of short circuit. A crowbar protection configuration is adopted against the over-voltage to have a controlled short circuit at the output and to force the intervention of the current-fold-back protection. The crowbar output silicon controlled rectifier (SCR) is individually heatsinked. In the design phase of the current-fold-back protection circuit, a high startup transient current that occurred on the +8.2 V module (Figure 6) had to be managed. This issue has been solved with a low-pass RC filter on the current-fold-back feedback control lines. Due to safety considerations and to simplify the production process, the same filter has been applied also to the +3.3 V high-power module.

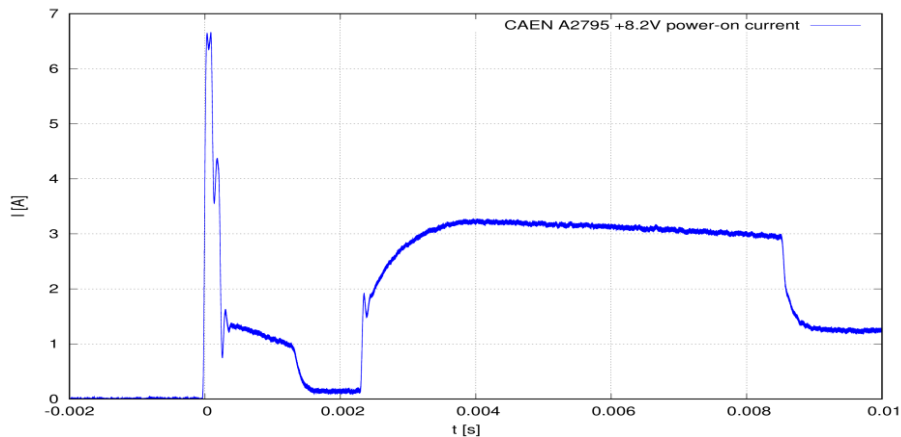

**Figure 6.** CAEN A2795 Module +8.2 V current power-up transient.

Due to the high current on the +8.2 V and +3.3 V outputs of the ICARUS analog readout crate, remote sensing of voltages is used to correctly regulate the voltage. In the design of the sensing circuit, great care has been taken to properly ground and decouple the various parts of the high-power supply module. Similar care has been taken in the design of the output filtering system, taking into account the parasitic RCL introduced by the supply cable. This has been vital to eliminate all the potential instabilities on the high-power supply module control loop.

Decoupling resistors on the high-power supply module are connected from the power rails to sense inputs in order to ensure a continuous voltage reference for the control loop even if no load is connected to the output. Their values exceed the supply cable wire resistance to avoid affecting the control loop when the output cable is connected. At the +8.2 V and +3.3 V outputs, sense wires are connected to the main supply rails via PTC resettable fuses. Their rating is lower than the ampacity of the sense wires, and their normal resistance value is ~0.1 $\Omega$ in order not to affect the sensing circuit. For the handling of the +8.2 V supply, each voltage rail has been split on two pins of the output connector to meet the current specifications of the contacts and the derating of the contacts' current capability, as required by the FNAL regulations. The supply cable is constructed with a single wire, where the two wires of each independent supply are twisted together. The +8.2 V wires are doubled and also twisted. The sensing wires are independently twisted and shielded, and all the power and sensing wires are globally shielded with a tinned copper mesh. The wire gauge (AWG) for the most current demanding power supplies is 14 AWG and has been determined for the estimated voltage drops for a 3 m long supply cable as listed in Table 2.

**Table 2.** ICARUS LVPS supply cable voltage drops.

| Vout [V] | Iout [A] | Sec [mm$^2$] | L[m] | Vdrop [V] |
|---|---|---|---|---|
| +8.2 | 14.0 | 5.0 | 3 | 0.34 |
| +3.3 | 7.4 | 2.5 | 3 | 0.35 |
| −5.0 | 1.2 | 2.5 | 3 | 0.06 |
| +12.0 | 0.7 | 2.5 | 3 | 0.03 |

The control board of the ICARUS LVPS that implements the power monitor server is based on the Aria G25 SMD module with an embedded Linux operating system. The communication with the on-board power monitor server is performed using the WebSocket protocol (RFC 6455) on port 4444. By default, the Ethernet is configured with DHCP enabled, and the network configuration can be changed by a touchscreen display or using a console debug connector (mini USB connector).

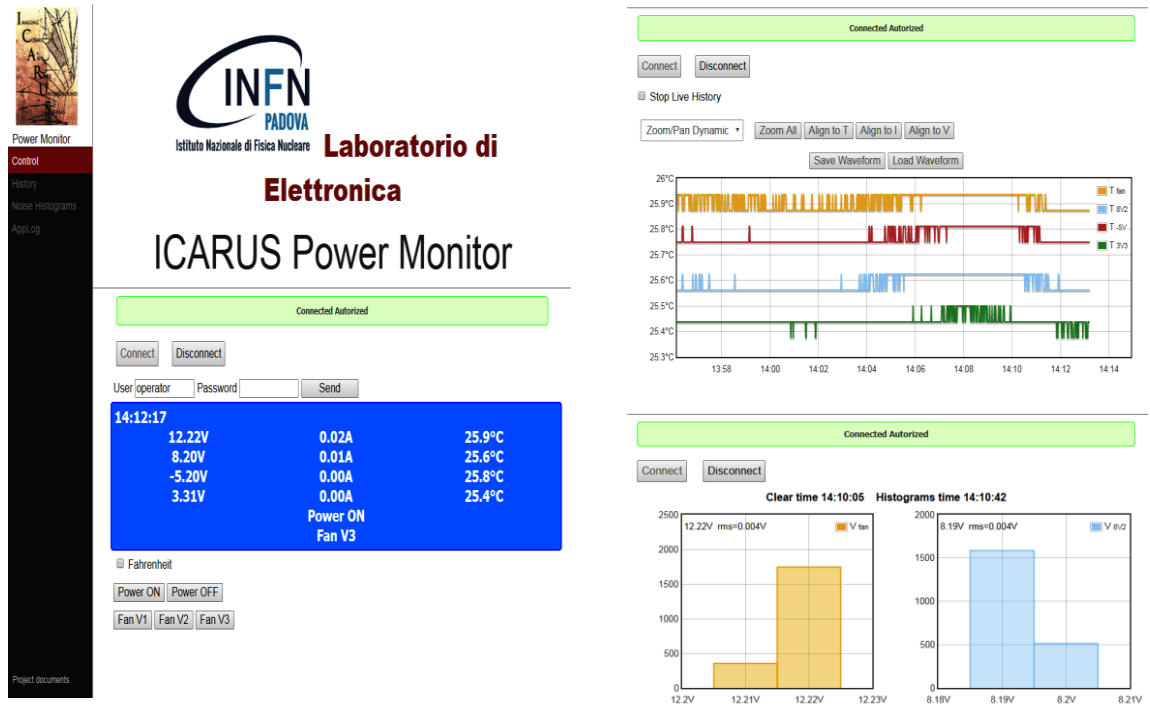

**Figure 7.** Screenshots showing the Web interface of ICARUS LVPS module. Initial page with the menu (left) and some monitors of the measured temperatures and distributions of the 12V and 8.2 V supply voltage (right).

The LVPS firmware allows for a remote control of some important functions, such as the ON/OFF of the power supply and the setting of the speed of the mini-crate's fans. Furthermore, the output voltages and the current values of the four power supply modules are read out at 100 Hz rate, while temperatures are read out at 1 Hz. Minimum, maximum and average values over a 1 s time window are stored in the internal memory, 1024 values deep, along with the measured temperature, assuring a registered history >15 minutes. Both the control and the monitoring functions are accessible via a web page, where the statistical information about the acquired parameters is also displayed and continuously updated (Figure 7).

## 5. Ultra-vacuum Feed-through flanges, Mini-crates and DBB's Cages

The new ultra-high vacuum feed-through flanges follow basically the same technology of the CF250 flanges already used at LNGS. The most relevant modifications concern both sides and borders of the flanges to allow for the installation of a mini-crate that houses nine A2795 boards directly inserted in the external connectors on the flanges, and of an internal card cage to support the new Decoupling and Biasing Board (DBB).

The study of the overall mechanical system supporting the TPC front-end electronics was divided into three phases including feasibility studies for a three-dimensional design of:
− The feed-through flange;
− The mini-crate containing nine A2795 electronic boards;
− The cage anchoring the new DBBs, connected to the flange on the internal side, and of the support of flat cables inside the ICARUS chimney.

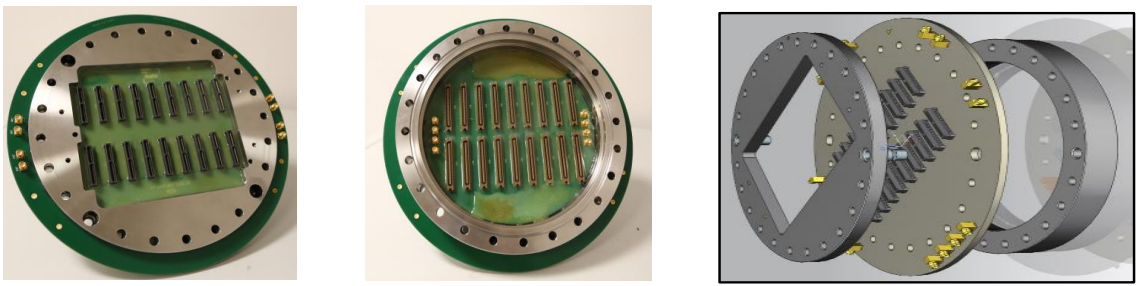

**Figure 8.** From left to right: external and internal side of the signal feed-through flange and the schematic of its basic structure. Components are glued together with a special process and vacuum/over-pressure tested by Criotec company (Turin, Italy).

The adopted solution led to the construction of the CF250 flange shown in Figure 8 by assembling the three main components:
− A stainless steel CF250 ring that interfaces the flange to the chimney;
− A multilayer PCB disk with connectors on both sides and two bias SMA connectors in the periphery of the disk;
− An external stainless steel disk with cutout for the base frame supporting the mini-crate.

The PCB disk is about 6 mm thick and is fabricated with 4 PCB layers with three interleaved copper layers to strengthen its structure. It holds both vacuum and overpressure. All the flanges have been tested over many pressure/vacuum cycles. They guarantee a vacuum leakage lower than $10^{-9}$ mbar·l/s and stand the differential overpressure of 1.1 bar, as required by safety rules. This proprietary design allows for obtaining a solid homogeneous disk without any external vias, that holds surface mounting connectors on both sides.

A major difference with respect to the previous design is the diameter of the PCB disk. In the present configuration the SMA connections, used for wire biasing and test pulsing, are hosted externally with respect to the stainless steel ring as shown in Figure 8 and Figure 9.

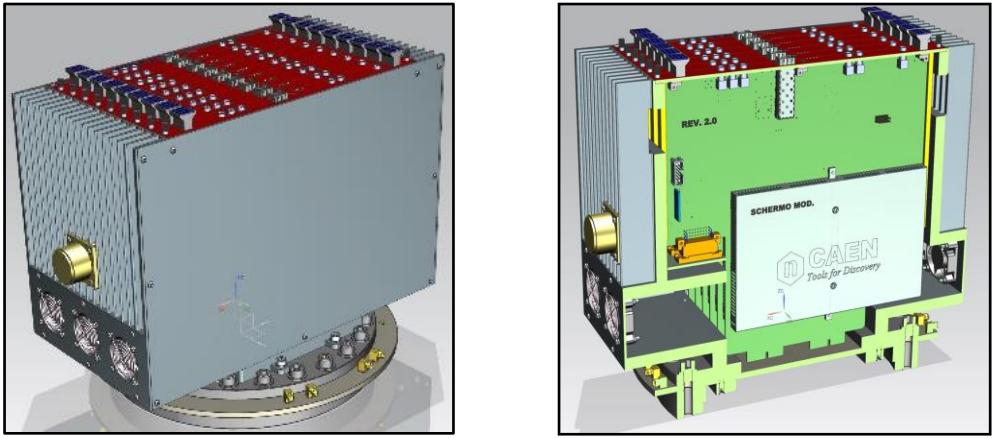

**Figure 9.** Rendering of the mini-crate with nine A2795 boards installed on the flange (left) and a cross section of the rendering, showing the internal layout of the mini-crate (right).

The mini-crate was designed and built in accordance with the request to make the structure light and at the same time rigid and compact (Figure 9). It must house the nine electronics boards, as well as the power and cooling distribution. It is composed of a base frame to interface with the CF 250 flange. The two cooling manifolds have been fixed to the base frame. The heat generated by the boards is channeled by the three small fans mounted on either side.

To guide the nine boards to their proper positions in the connectors welded on the external side of the flange, nylon guides were fixed on the finned profiles screwed to the support plate (Figure 10). The finned profiles also served as an additional body for exchanging the heat generated by the boards. They have openings in the upper part for the exit of cooling air. The new mini-crate installed on a feed-through flange is shown in Figure 11.

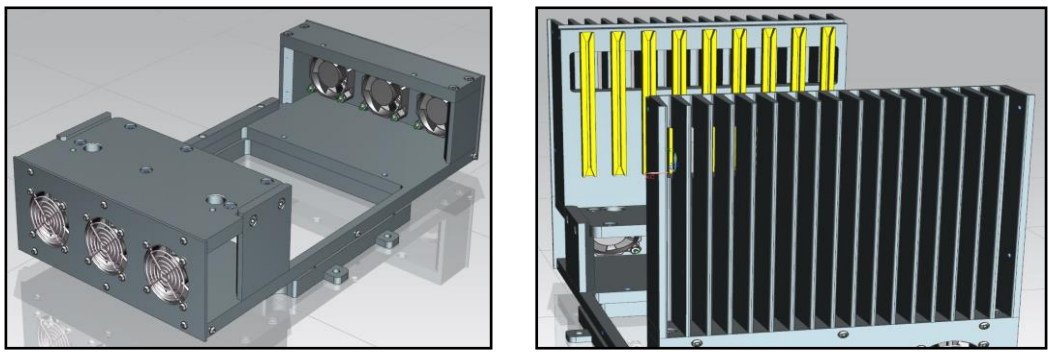

**Figure 10.** Detail of the base frame (left) and card guides fixed on the finned heat exchangers (right). In the guide in the back the window for venting cooling air is visible.

The interface between the signals from wires and the external electronics modules in the mini-crate is provided by the nine DBB inserted in the internal connectors of the flange, mechanically constrained in a metal cage. Supports and combs that serve to guide the DBBs and lock them to the inner connectors of the flange have been designed to avoid any possible mechanical stress to the connectors due to the considerable stiffness and weight of the flat cables. Two L-shaped stainless steel plates were used to support and block the DBBs (Figure 12). Clamps have been designed to give the right curvature to the flat cables to account for their low flexibility and to support their weight.

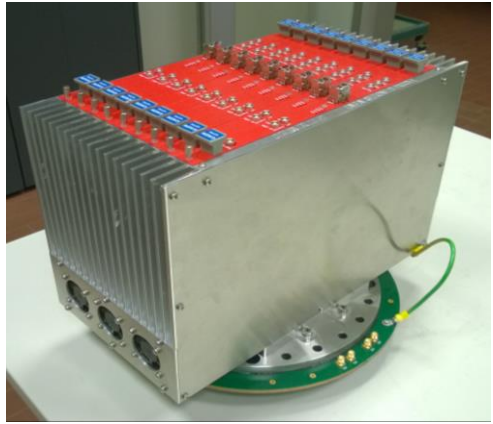

**Figure 11.** A mini-crate populated by the nine A2795 boards installed on a feed-through flange. The strap connecting the ground of the feed-through flange to the mini-crate is also shown.

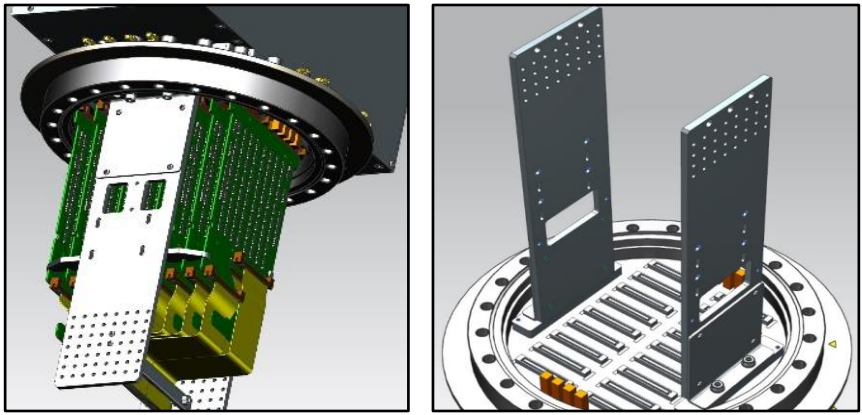

**Figure 12.** Rendering of the flange fully equipped with the DBBs held in place (left) and of the internal view of the flange with the side support of the DBB cage (right).

## 6. Decoupling and Biasing Boards (DBB)

The connection of the TPC wires, the high voltage bias decoupling, and the scheme to pulse the Induction 2 and Collection wires at their far side (opposite side with respect to the signal connector) shown in Figure 13 are described in [5].

The ICARUS-T600 TPC wire signals are fed into the front-end amplifiers by means of DBBs (Figure 12) designed and built by INFN Padova (Italy) staff. Due to the constraints from both the TPC and flange configurations, the DBB has been designed to host two galvanically isolated 32 channel banks. This solution allowed for biasing each bank independently without any stray current (i.e. leakage currents) in between, preventing any related noise contribution.

The 53248 ICARUS TPC wires are connected to 1664 32-channel flat cables and served by 856 DBBs on 96 flanges, 9 DBB cards (576 channels) being hosted on each flange. The DBB is a crucial RC link in the ICARUS signal extraction chain while providing bias voltage for the wires. It has to work in argon gas and operate up to 300 Vdc, without adding parallel noise to the wire signals. Moreover, any possible failure on a single channel must not influence and jeopardize the operation of the other channels. In addition to the traditional circuit design best practices, it has been mandatory to study and prevent all the potential failure conditions because the DBBs are installed inside the cryostats in unreachable positions without any access for maintenance. This required some redundancy in the circuitry and in-depth *ex ante* test protocols before their installation in the detector.

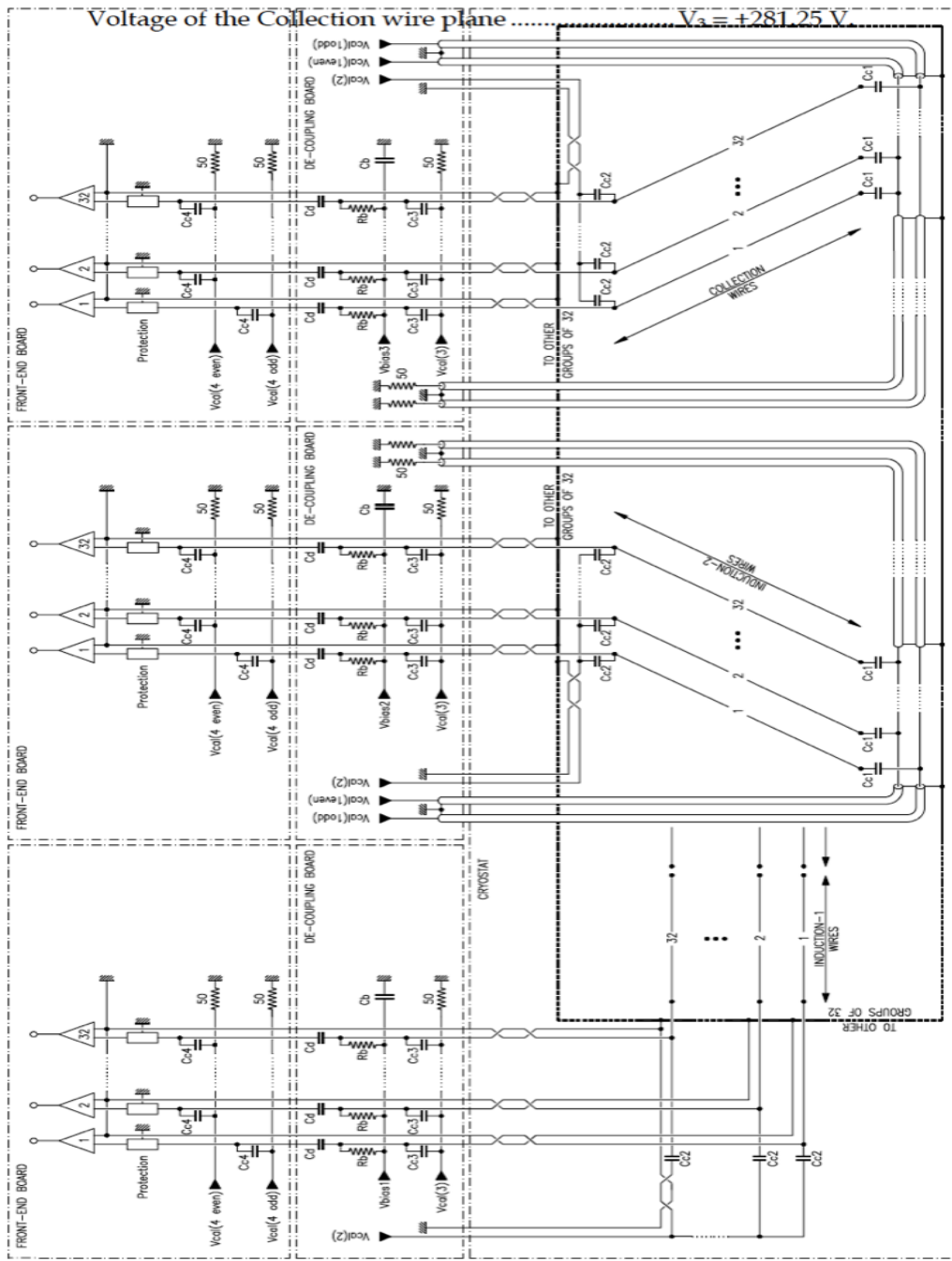

**Figure 13.** Read-out, high voltage decoupling and calibration scheme. All GND symbols are intended to be detector ground.

– 13 –

DBB were realized using a double side PCB without internal layers (Figure 14). All the HV signals brought by the TPC wires run on the upper layer whilst all the low voltage capacitively decoupled signals run on the bottom one. A star network configuration is implemented in the circuit grounding design. A guard ring guarantees galvanic isolation between the two 32-channel banks thus avoiding any leakage current possibly due to different biasing HV applied to each bank.

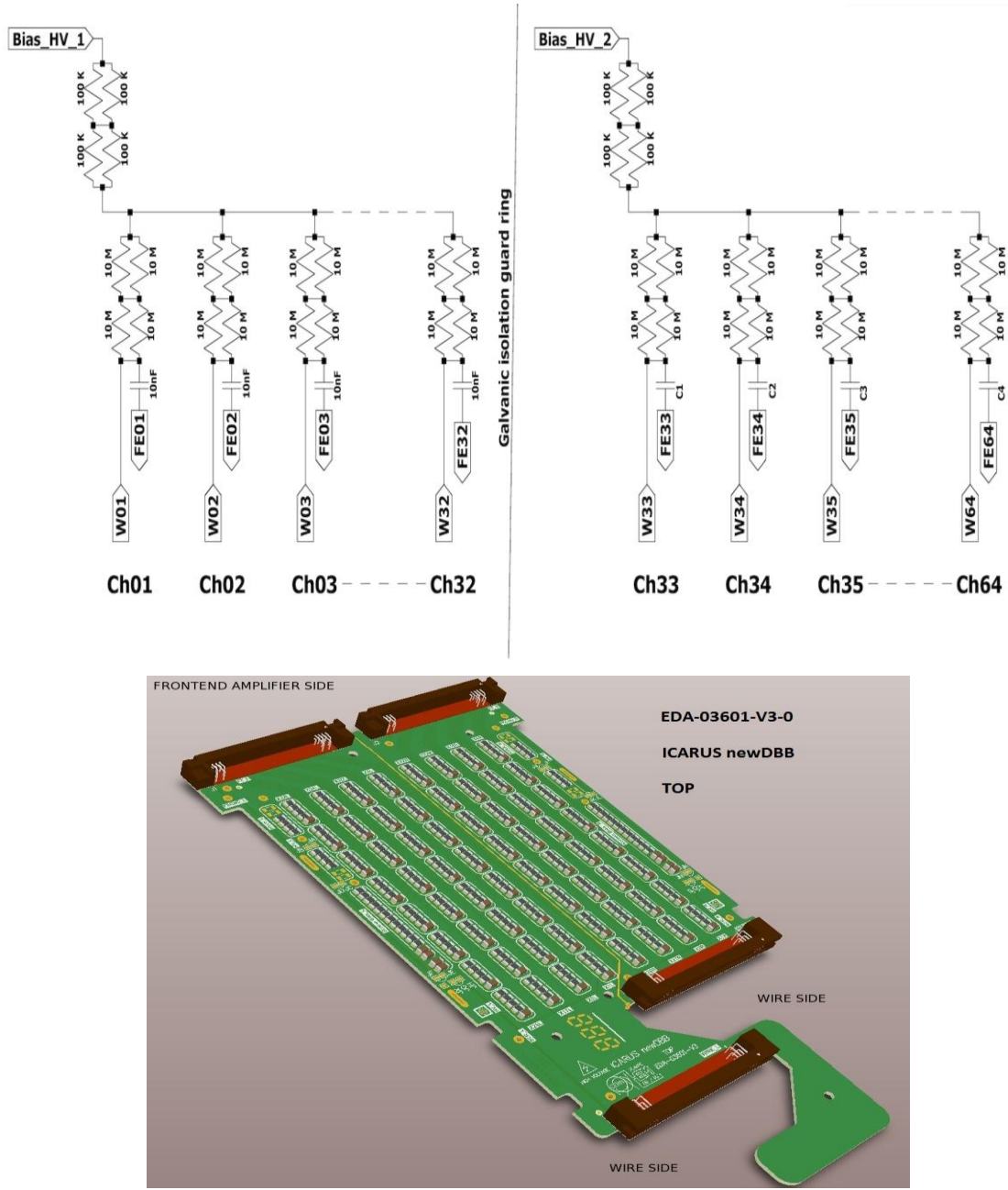

**Figure 14.** Schematics (top) and rendering (bottom) of the DBB.

Only 1206 size ceramic substrate SMD resistors and Multi-Layer Ceramic Capacitors are used on the DBB, which are the largest size compatible with the DBB allowed dimensions. All the resistors and capacitors are mounted with the long side of the 1206 package parallel to the connector side of the PCBs to minimize breakage risks of the ceramic substrate components due to unwanted bending of the PCB. To minimize the risk of failure, the 10 MΩ biasing resistors are realized as the series of two parallels of two 10 MΩ resistors, making highly improbable the complete failure of the biasing of a channel. The failure of one resistor will not change the biasing voltage of the wire, but only will slightly change the impedance of the wire to the ground without affecting the amplitude of the signal. The 10 nF decoupling capacitor adopted the NP0-C0G technology which guarantees no offset in the capacitance value and no noise due to the piezoelectric effect typical of the X7R technology previously considered. A high quality HV input hookup cable has been selected to connect each 32-channel bank of the DBB to the HV input. The custom connectors have been developed by the KEL company.

All the 890 produced DBBs have been tested following two different protocols:
− Static test at bench to guarantee the integrity and functionality of the DBBs. The test has been performed by means of spring-loaded probes which measure the equivalent resistance/capacitance of subsections of the circuit (5 resistance and 2 capacitance measurements). Only 3 capacitors out of 71200 and 3 resistors out of 234960 did not pass the test. All the DBBs with bad components were discarded. All the 864 DBBs actually installed in the T600 passed the test without any hardware modification.
− HV test at bench to exclude unwanted current generated by HV biasing. These currents may be related to leakage through the capacitor's dielectric or stray surface current between channels. Ten DBBs showing a suspicious behavior have been discarded. Enough DBB spares are anyway available.

## 7. Installation of the TPC read-out electronics on T600 detector and first tests

The installation of the DBBs in the inner side of the flanges has been performed in a clean area next to the T600 detector at FNAL (Figure 15). The leakage current was measured at the nominal HV on each assembled feed-through flange: no current excess above the value I = 20 nA expected for the decoupling capacitors was found.

Feed-through flanges are installed on the top cross of the chimneys of the T600. The flat signal cables from the TPC wires, grouped in bundles of 18 cables and pre-routed to the 96 chimneys, are connected to the DBBs. Moreover, the two coaxial cables for the TPC wire biasing are connected to the SMA feed-through on the inner side of the flange in each chimney, as well as the four cables for the external test pulse signals. The function of these four cables is the same as described in Ref. [5].

Electrical continuity tests have been performed on each TPC wire before closing/tightening it to its chimney. A custom-designed test box (Figure 16 left) injected a test pulse on to the test capacitors present at the far end of TPC wires. Horizontal wires that terminate on the ends of the TPC do not have test capacitors. For these, signals are induced by pulsing other wire planes. The induced signals on the corresponding signal cables are amplified and displayed on an oscilloscope, as shown in Figure 16 (right). The design of the test box allows collaborators to quickly test all of the TPC wires for continuity. The first round of tests was done observing signals at the end of each signal cable and then later, after the DBBs and flanges had been installed, on the warm side of the flange. Several problems were found and corrected before closing of the TPC.

Mechanical stability of the flange system, a necessary condition for electrical continuity, has been checked by measuring the overall equivalent impedance $Z_{Th}$ between the HV bias input and the front-end amplifier outputs for each DBB chain (flange to DBB and the flat cable bundle to the TPC wires). ZTh has proven to be a very stable and reproducible parameter yielding to a very sensitive test. The test has been conducted after each manual intervention on the flange and after extended standby periods (1-2 months).

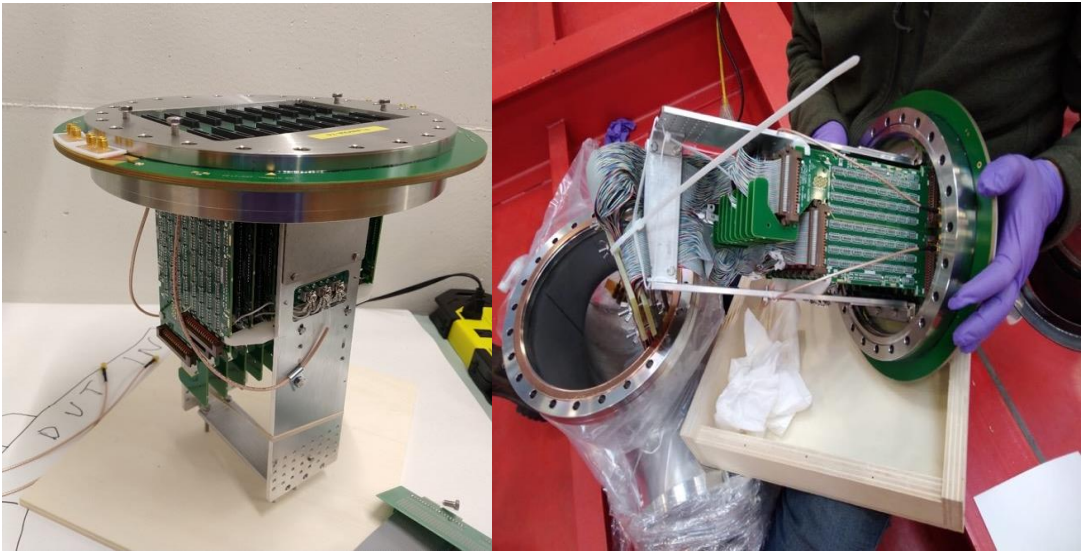

**Figure 15.** An assembled feed-through with nine DBB and the biasing cables (left). Installation of a feed-through flange on the top cross of a chimney (right). The bundle of the 18 flat signal cables and the two brown coaxial biasing cables are visible.

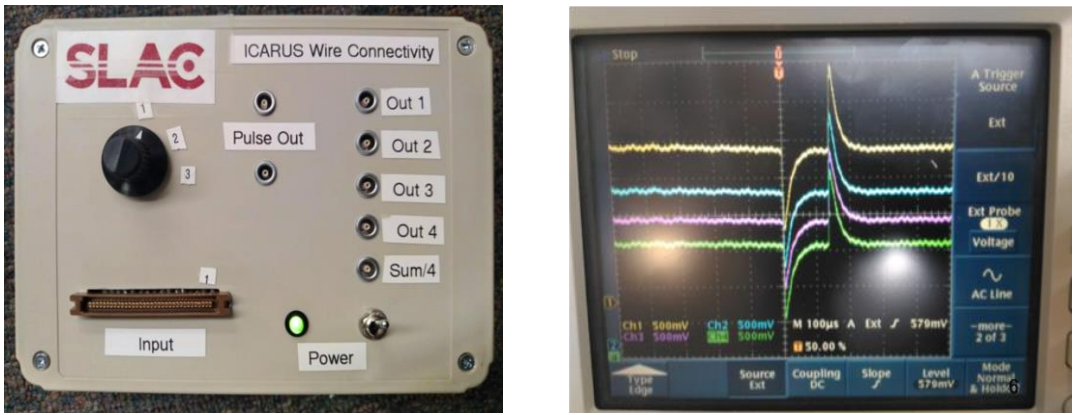

**Figure 16.** Custom-designed test box (left). Induced signals displayed on the oscilloscope (right).

Mini-crates are then installed on the feed-through flanges and populated with the A2795 boards already equipped with the custom-designed amplifiers previously tested at bench. The associated LVPS modules located in small racks between the mini-crates were turned on and their functionality tested (Figure 17).

Each mini-crate needs to be initialized to be synchronous with the full DAQ system to provide a consistent time stamp for all the collected events. This function is performed by a 10 MHz TTL waveform that is distributed (TT-Link) as shown in Figure 18. Among other control signals, the TT-Link command, generated by a SPEXI module (INCAA computers, The Netherlands) provides the start signal to read out all the mini-crates. Special fan-out units, built at Colorado State University (USA), distribute the TT-Link signals equalizing the delay for all the mini-crates as shown in the figure.

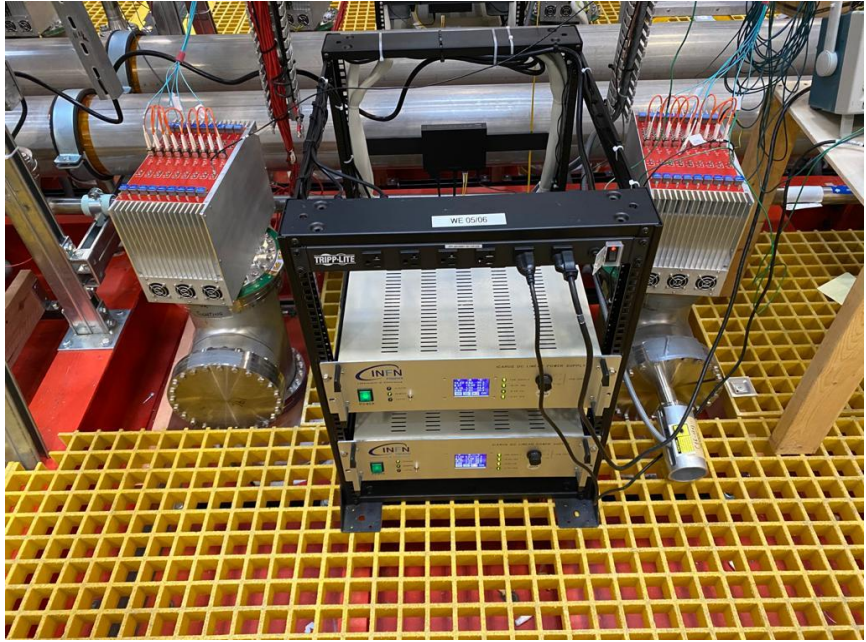

**Figure 17.** Two LVPS modules powering the two adjacent mini-crates populated with nine A2795 boards, serving 576 wires each.

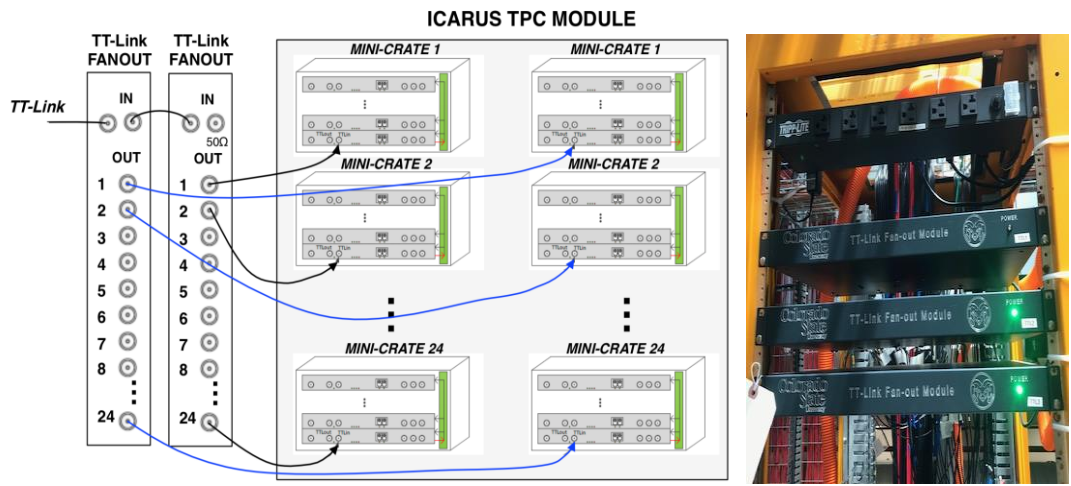

**Figure 18.** Layout of the TT-Link distribution with delay equalization (left) and the corresponding two switched on fan-out (green light on) modules on top of the rack serving one ICARUS TPC module (right).

An initial TPC vertical slice test including the full read-out electronics has been performed on a few chimneys on top of the ICARUS-T600 by injecting test pulses and reading out the wire signals via the A2795 to evaluate the overall performance (Figure 19). A PC with one A3818 CAEN board installed was used to receive data through the optical links. Besides the huge charge injected, compared to ~2 fC of a m.i.p., this initial test confirmed the functionality of the full TPC electronics. This test was carried out with the TPC at room temperature with the wires in air.

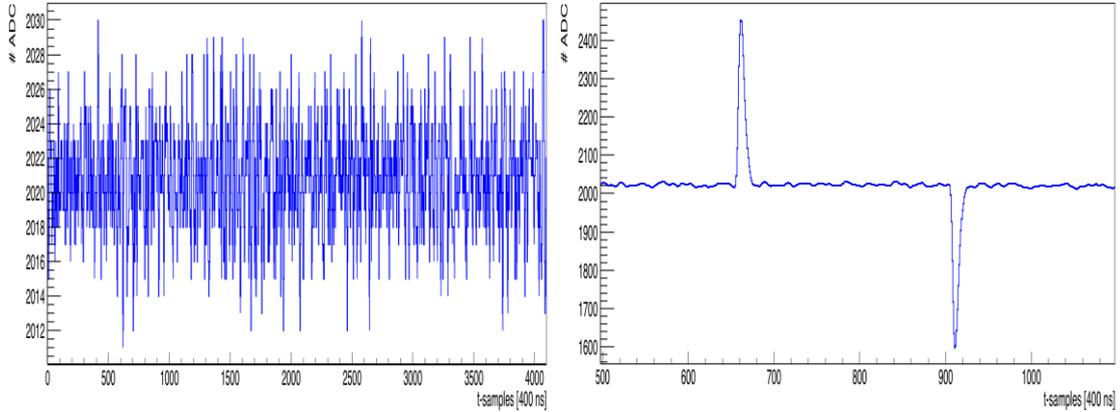

**Figure 19.** Typical baseline noise, RMS = 3.1 ADC counts without LAr in the TPC, (left, 1 ADC count corresponds to ~550 electrons with the present 12 bit ADC) and shape of the injected test pulse as detected on one Collection wire (right).

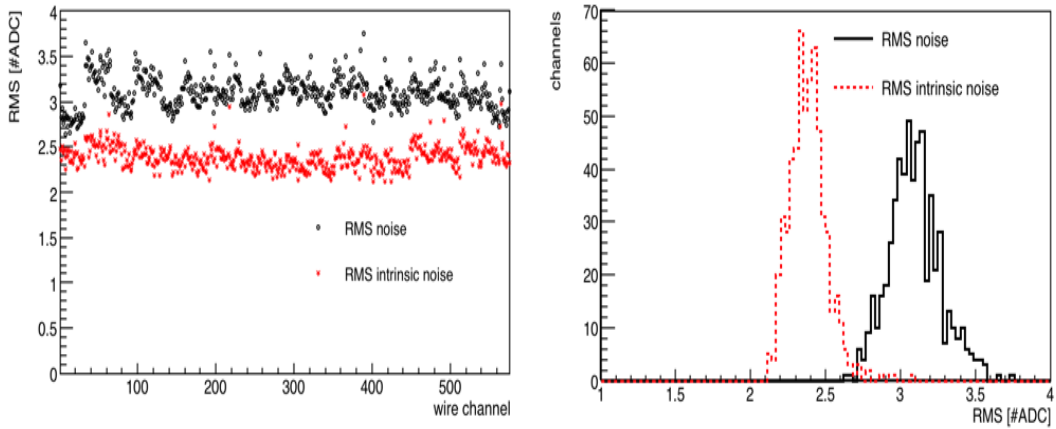

**Figure 20.** The measured RMS noise in ADC counts for the 576 channels in a mini-crate as determined by averaging over 100 events (left) and the corresponding overall 576 channels distribution (right). In each board, the first 32 channels refer to Induction-2 wires and the second 32 to the Collection wires. Black points (solid line) refer to the measured values, while the red ones (dashed line) to the intrinsic noise after the coherent component removal.

The noise level has been measured collecting random triggers with no pulse signal injected. In Figure 20, the RMS noise detected on each wire is shown for all the 576 channels of a mini-crate, each of the 9 boards serving 32 Induction-2 wires (channels 0-31) and 32 Collection wires (channels 32-63). Without optimizing grounding conditions, a satisfactory average noise level has been measured without the wires submerged in LAr: ~1700 electrons compared to the ~1200 electrons design value in Ref. [8]. A common noise structure, related to the basic 32 wire modularity of the electronic chain, is still observed as already detected in the test bench at CERN [8]. The removal of this coherent noise component reduces the average noise level to 2.4 ADC counts corresponding to ~1300 electrons to be compared with the corresponding 1100 electrons as observed in the test bench.

Eventually all 53248 channels have been tested and noise level measured. The noise level for the horizontal wires (Induction-1) is slightly higher due to the longer wire and cable length. The test has been done also with the PMTs turned on and no effects on noise were observed. PMT system has been described in Ref. [4].

## 8. Conclusions

A new full TPC read-out electronics together with a new wire biasing and interconnection system have been realized for the ICARUS-T600 operation at FNAL. It includes a new design of the analogue front-end, digitizer and serial bus architecture with optical links for Gb/s data transmission. A new compact set-up has been produced based on a custom signal feed-through flange acting as the electronic back-plane and a dedicated mini-crate hosting the front-end boards. New decoupling boards have been directly installed in the internal side of the flanges. New low voltage power supplies characterized by an extremely low noise to minimize any external influence on the TPC read-out performance have been realized. All the electronics serving the 53248 channels have been installed and tested on the T600 detector at FNAL. With the present detector grounding conditions far from optimal, an initial satisfactory noise level has been measured. The installation of the full TPC electronic chain has been successfully accomplished with the cables deployed. A systematic study of the TPC electronics performance will follow the LAr filling of the ICARUS-T600 detector, aiming to improve noise conditions by carefully checking all possible ground loops or any other issues.

**Acknowledgments**

This work was funded by INFN in the framework of WA104/NP01 program finalized to the overhauling of ICARUS detector in view of its operation on SBN at FNAL, and by the European Union's Horizon 2020 research and innovation program under the Marie Sklodowska-Curie grant agreement No 822185, and by the U.S. Department of Energy, Office of Science, Offices of High Energy Physics and Nuclear Physics. Fermilab is operated by Fermi Research Alliance, LLC under Contract No. DE-AC02-07CH11359 with the United States Department of Energy. The A2795 board was designed, engineered, and built by the CAEN company, in collaboration with ICARUS team and Electronics Service facilities of INFN, Padova. The outstanding contributions to the construction and installation of the full ICARUS TPC electronic chain of the Electronics Service, Mechanical Design, and Mechanics facilities of INFN Padova are also acknowledged. As is the excellent support of the Fermilab technical staff.